\documentclass[prb,twocolumn,preprintnumbers,amsmath,amssymb]{revtex4}
\usepackage{graphicx}

\begin{document}

 \title{ \bf  Quantum phase transitions, Excitonic supersolid  and its
  detections in electron-hole bilayer systems   }
\author{ \bf  Jinwu Ye  }
\affiliation{ Department of Physics, The Pennsylvania State
University, University Park, PA, 16802 }
\date{\today}

\begin{abstract}
    We construct a quantum Ginsburg-Landau theory to
    study the quantum phases and transitions in electron hole bilayer system.
    We propose that in the dilute limit as distance is increased,
    there is a first order transition from the excitonic superfluid (ESF ) to the excitonic
    supersolid (ESS) driven by the collapsing of a roton minimum, then a 2nd order transition from the ESS to  excitonic
    normal solid.
    We show the latter transition
    is in the same universality class of superfluid to Mott
    transition in a rigid lattice.  We then study novel elementary low energy excitations inside
    the ESS. We find that there are two "supersolidon " longitudinal
    modes ( one upper  branch and one lower branch ) inside the ESS,
    while the transverse mode in the ESS stays the same as that inside a ENS. We also work out
    various experimental signatures of these novel elementary excitations
    by evaluating the Debye-Waller factor, density-density correlation, specific heat and vortex
    -vertex interactions. For the meta-stable supersolid generated by photon pumping, we show that the angle resolved
    spectrum is dominated by the macroscopic super-radiance from its superfluid component, even it is
    just a very small percentage of the the whole system. This fact
    can be used to detect the metastable ESS state generated by photon pumping by a power spectrum experiment
    easily and without any ambiguity.
\end{abstract}

\maketitle

\section{ Introduction }

    Recently, degenerate exciton systems have been produced by different experimental groups
    with different methods  in
    quasi-two-dimensional semiconductor $ GaAs/AlGaAs $ coupled quantum wells structure
    \cite{butov,snoke,bell,field1,field2}. There are two different
    ways to generate the excitons. One is through photon pumping, then applying an electric field
    along $ z $ direction to confine electrons in one  quantum well and holes in another quantum well
    \cite{butov,snoke,bell}. We call these kind of excitons as
    photon generated excitons.
    The second is through gate voltage \cite{field1,field2}.
    We call these kind of excitons as gate voltage generated excitons.
    When the distance between the two quantum wells is sufficiently small,
    an electron in one well and a hole in the other
    well could pair to form an exciton which behaves as a boson at long distance.
    Now it was widely believed that the electron-hole bilayer (EHBL) is a very
    promising system to observe Bose-Einstein condensation (BEC) of excitons.
For simplicity, we only discuss balanced el-hole bilayers where the
electrons and holes have the same density $ n_{e}=n_{h} $, but may
have different masses $ m_{e} \neq m_{h} $.
 The case with different masses and with different density is
    also very interesting and will be investigated in separate publications.

There are two important dimensionless parameters in the EHBL. One is
the dimensionless distance $ \gamma= d/a_{B} $ (  $ a_{B} $ is the
Bohr radius )  between the two layers. Another is $ r_{s} $ which is
the ratio of the kinetic energy over the potential energy in a
single layer. The $ r_{s} a_{B} $ is the typical interparticle
distance in a single layer.  It is easy to see that the ratio of
intralayer Coulomb $ V_{11} $ over the interlayer Coulomb $ V_{12} $
    interactions $  \alpha =  V_{11}/V_{12} = d/ r_{s} a_{B} $.
    So when the interlayer Coulomb interaction dominates  $ \alpha < 1
    $, the EHBL is expected to exhibit the superfluid of excitons.
If the density of excitons is sufficiently low (  large $ r_{s} $ ),
then the system is in a weakly coupled Wigner solid  state at very
     large distance and become an BEC excitonic superfluid ( ESF ) at short
     distance ( Fig.1). An interesting problem is how the system evolves
     from the BEC ESF to the weakly coupled Wigner solid as the distance
     increases.
    If an exciton is already formed, its kinetic energy $ K \sim \frac{ \hbar^{2} }{ m^{*} (r_{s}
    a_{B} )^{2} } $, its potential energy $ P \sim \frac{ e^{2}
    d^{2} }{ \epsilon ( r_{s} a_{B} )^{3} } $. When $ K < P $,
    namely, $ \sqrt{r_{s}} <  d/a_{B} $, the EHBL could favor a excitonic ( or dipolar) normal solid (ENS) state.
    As argued above, when $ d/a_{B} < r_{s} $, the EHBL is in a ESF state.  So
    in the intermediate distance $ \sqrt{r_{s}} < d/a_{B} < r_{s} $ \cite{hand},
    the system may favor a excitonic ( or dipolar ) supersolid (ESS) state.
    When $ d/a_{B} > r_{s}  $, it will become the excitonic normal solid (ENS)
    due to the long range dipole-dipole $ 1/r^{3} $ repulsive interactions ( Fig.1 ).
    A hole system with  $ r_{s} \sim 20-30 $ was already realized \cite{hole}.
    If $ r_{s} \sim 20 \gg 1 $ limit, there is a broad distance regime $ 4.5 < d/a_{B} < 20
    $, the system could be in the ESS state.
    It becomes feasible to experimentally explore all
    the possible phases and phase transitions in the EHBL in the near future.

    It was originally pointed out in \cite{ches} that for any supersolid to exist, there
    must be quantum fluctuations generated vacancies even in the ground state of a solid at $ T=0 $.
    Obviously, the first candidate to search for supersolid is near
    the phase boundary of superfluid $ ^{4} He $ and solid $ ^{4} He
    $. The authors in \cite{kim} suggested that the
    supersolid state leads to the
    non-classical rotational inertia (NCRI) observed in the torsional
    oscillator experiment, many other experiments such as
    neutron scattering, X-ray diffraction, mass flow, heat capacity,
    acoustic sound attenuation and so on are needed to
     uniquely distinguish the supersolid from other enormous number of much less
     interesting possibilities. A phenomenological quantum Ginzburg-Landau theory \cite{qgl} was
     developed to make predictions on signatures of these
     experiments if the supersolid indeed exists in Helium 4.
     Unfortunately, so far, all these experiments came as
     negative. It becomes interesting to see if  a supersolid state can exist
    in other systems which also have both superfluid state and solid state.
    In this paper, we will point out that the excitons in EHBL maybe a
    very promising experimental system to observe the excitonic supersolid (ESS) near the
    phase boundary between the ESF and the ENS.
    Because the exciton's mass
    $ m_{ex}=0.37 m_{e} $ is much smaller than that of $ ^{4}He $,
    so the zero point quantum fluctuations in the ENS in EHBL is even much
    larger than those in solid Helium 4. I will point out a new
    mechanism which is absent in $ ^{4}He $ to generate repulsive excitonic vacancies in
    the ENS which leads to the intermediate excitonic supersolid (ESS) phase  in the dilute limit in the
    EHBL.  Then we construct a quantum Ginzburg-Landau theory to study all the
    phases and quantum phase transitions in Fig.1.
    It is instructive to compare the EHBL system with the
    pseudo-spin sector in the bilayer quantum Hall system  ( BLQH ) at total
    filling factor $ \nu_{T}=1 $.
    Although the ESF in Fig.1 shares some common properties with the
    corresponding ESF in the BLQH \cite{psdw,imb,cbtwo},  due to different symmetries of the two
    systems, translational symmetry breaking states are very
    different.
    For example, ESS and ENS phases
    which are the focus of this paper have no analogies in the
    BLQH.  These crucial differences will be explicitly studied in the following.
    Furthermore, in sharp contrast to BLQH which is a stable system,
    for the photon generated excitons, the excitonic phases in Fig.1 are just meta-stable states which
    will eventually decay by emitting lights.  We will show that
    the angle resolved power spectrum from the internal photon can
    detect the ESS unanimously if it indeed exists.
    In fact, the  characteristics of emitted photons is a very natural, feasible and unambiguous internal probe of
    all the three phases ESF, ESS and the ENS in Fig.1.

      The paper is organized as follows. In sec. II, we derive the
   quantum Ginsburg-Landau action to describe the transition from
   the ESF to the ENS driven by the collapsing of a roton minimum.
   Then In Sec.III, we will argue that in general there should be a
   ESS state intervening between the ESF and the ENS. In Sec.IV,
   by renormalization group analysis, we study the
   universality class of zero temperature quantum phase transition
   from ENS to ESS driven by the
   distance. In Sec.V, we work  out the elementary low energy
   excitations inside the supersolids. Then in the following
   sections, we study the experimental signatures of these low
   energy excitations by calculating the Debye-Waller factor in the X-ray
   scattering from the ESS in sec. VI, the density-density
   correlation function in the ESS in sec.VII.
   In Sec.VIII, we study the specific heat in the ESS.  In Sec. IX, by
   performing a duality transformation to the vortex
   representation, we will study the vortices in the ESS.
   In Sec.X, we will present the photon emission pattern from
   the ESS formed by photon pumping generated excitons.
   Finally, we reach conclusions in Sec.XI.



\section{ The zero temperature transition from ESF to ENS driven by
the distance }

     If $ c_{1} ( c_{2} ) $ is the electron annihilation operator in top ( bottom )
    layer, then $ h^{\dagger}_{2} = c_{2} $ is the hole creation
    operator in the bottom layer.
    The order parameter for the ESF is the  $ p-h $ pairing $ \psi (
    \vec{x},\tau )= \langle c^{\dagger}_{1} c_{2} \rangle= \langle c^{\dagger}_{1} h^{\dagger}_{2}\rangle =
    \sqrt{ \bar{\rho} + \delta \rho } e^{ i \theta( \vec{x},\tau ) }  $. The effective action inside
    the ESF is essentially the same as that in the pseudo-spin
    channel in BLQH \cite{psdw}:
\begin{equation}
 {\cal L}[ \delta \rho, \theta ]= i \delta \rho  \partial_{\tau} \theta +
          \frac{ 1 }{2}  \rho_{d} ( \nabla \theta )^{2}
          + \frac{1}{2}\delta \rho V_{d} (\vec{q} ) \delta  \rho
\label{symm}
\end{equation}

    In the ESF state, it is convenient to integrate out
    $ \delta \rho $ in favor of the phase field $ \theta $ to get
    the phase representation:
\begin{equation}
    {\cal L}[ \theta ] = \frac{1}{ 2 V_{d}( \vec{q} ) } ( \partial_{\tau} \theta )^{2} +
     \rho_{d} ( \nabla \theta )^{2}
\label{phase}
\end{equation}
     where the dispersion relation of the Goldstone modes including higher
     orders of momentum can be extracted:
\begin{equation}
     \omega^{2} = [2 \rho_{d}  V_{d}( \vec{q} ) ] q^{2}
\label{disper}
\end{equation}

      In the long wavelength limit,
      $ V_{d}( \vec{q} \rightarrow 0 )
      \rightarrow  c $ ( $ c $ is a constant ) ( Fig.1 ) leads to a capacitive term for the
      density fluctuation.
      The QMC calculations  \cite{eh1,eh2} indeed find that there is a roton minimum
      in the dispersion relation ( Fig.1).

    Because the original instability
    comes from the density-density interaction $ V_{d}( \vec{q} ) $ ,
    it is convenient to integrate out the phase field in favor of the
    density operator in the original action Eqn.\ref{symm}.
    Neglecting the vortex excitations in $ \theta $ and integrating out the $ \theta $ in
    Eqn.\ref{symm} leads to:
\begin{equation}
 {\cal L}[ \delta \rho ]=
    \frac{1}{2} \delta \rho(-\vec{q},-\omega_{n} ) [ \frac{ \omega^{2}_{n} }{ 2  \rho_{d} q^{2}}
    + V_{d} (\vec{q} ) ] \delta \rho(\vec{q},\omega_{n} )
\label{density}
\end{equation}
    where we can identify the dynamic density-density correlation
    function:
\begin{equation}
    S(\vec{q},\omega_{n} )
    = \langle \delta \rho(-\vec{q},-\omega_{n} ) \delta
    \rho(\vec{q},\omega_{n} ) \rangle =  \frac{ 2 \rho_{d} q^{2} }{ \omega^{2}_{n} +
   v^{2}(q) q^{2} }
\label{dd}
\end{equation}
    where $ v^{2}(q)= 2 \rho_{d} V_{d} (q) $ is the spin wave velocity defined in Eqn.\ref{disper}.

    From the pole of the dynamic density-density correlation
    function, we can identify the speed of sound
    wave which is exactly the same as the spin wave velocity.
    This should not be too surprising. As shown in liquid $ ^{4}He $,
    the speed of sound is exactly the same as the phonon
    velocity. Here, in the context of excitonic superfluid, we
    explicitly prove that the sound speed is indeed the same as the spin
    wave velocity.

    From the analytical continuation
    $ i \omega_{n} \rightarrow \omega + i \delta $ in Eqn.\ref{dd}, we can identify the dynamic
    structure factor: $ S( \vec{q},\omega ) = S(q) \delta ( \omega
    -v(q)q ) $ where $ S(q)= \rho_{d} q \pi/v(q) $ is the equal time
    density correlation function shown in Fig.1. As $ q
    \rightarrow 0, S(q) \rightarrow q $.
    The {\sl Feymann relation }  in the ESF which relates the dispersion relation to the equal
    time structure factor is
\begin{equation}
     \omega (q) =  \frac{ \rho_{d} \pi  q^{2} }{  S(q) }
\end{equation}
    which takes exactly the same form as
    the Feymann relation in superfluid $ ^{4} He $.
    Obviously, the $ V_{d}(q) $ in the Fig.1b leads to the roton dispersion $ \omega^{2}= q^{2} V_{d}(q)
    $ in the Fig.1a.

    Because the instability happens near $ q=q_{0} $ instead of $ q=0 $, so the
    transition in Fig.1 is {\em not} driven by vortex unbinding
    transitions like in 3D XY model, so the vortices remain tightly
    bound near the transition. So integrating out
    the vortex excitations in $ \theta $ will only generate
    {\em weak} interactions among the density $ \delta \rho $:
\begin{eqnarray}
 {\cal L}[ \delta \rho ] & = &
    \frac{1}{2} \delta \rho(-\vec{q},-\omega ) [ \frac{  \omega^{2}_{n} }{ 2 \rho_{d} q^{2}}
    + V_{d} (\vec{q} ) ] \delta \rho(\vec{q},\omega )  \nonumber  \\
    & - & w ( \delta \rho )^{3}+  u ( \delta \rho )^{4} + \cdots
\label{densityint}
\end{eqnarray}
    where the momentum and frequency conservation in the quartic and
    sixth order  term is assumed.

      Expanding $ V_{d}(q) $ near the roton minimum $ q_{0} $
     leads to the quantum Ginsburg-Landau action to describe the ESF to the ENS transition:
\begin{equation}
 {\cal L}[ \delta \rho ]  =
    \frac{1}{2} \delta \rho [ A_{\rho} \omega^{2}
    + r+ c( q^{2}-q^{2}_{0} )^{2} ] \delta \rho
     -w ( \delta \rho )^{3} + u ( \delta \rho )^{4} + \cdots
\label{densitymin}
\end{equation}
    where  $  A_{\rho} \sim \frac{ 1 }{ 2 \rho_{d} q_{0}^{2}} $ which
    is non-critical across the transition.
     In sharp contrast to the ESF to the pseudo-spin density wave (PSDW)
     transition in BLQH \cite{psdw}, because of the lack of $ Z_{2} $
     exchange symmetry between the two layers in EHBL, there is a cubic term in
     Eqn.\ref{densitymin}. It was explicitly shown in \cite{loff} that in the presence  of
     both the cubic and the quartic terms, the favorable lattice is a triangular
     lattice instead of a square lattice in PSDW \cite{psdw}.
     The generic transition driven by the collapsing of roton minimum is from ESF to ENS instead of
     from the ESF to the excitonic supersolid ( ESS ). In the ESF, $ r \rangle 0, \langle \psi \rangle \neq 0,  \langle \delta \rho \rangle =0 $,
     In the ENS, $ r < 0, \langle \psi \rangle = 0, \langle \delta \rho \rangle = \sum^{\prime}_{\vec{G}}  n_{\vec{G}} e^{i \vec{G} \cdot
      \vec{x} } $ where $ \vec{G} $ is the 6
      shortest reciprocal lattice vector of a triangular lattice ( Fig.1 ).

\begin{figure}
\includegraphics[width=8cm]{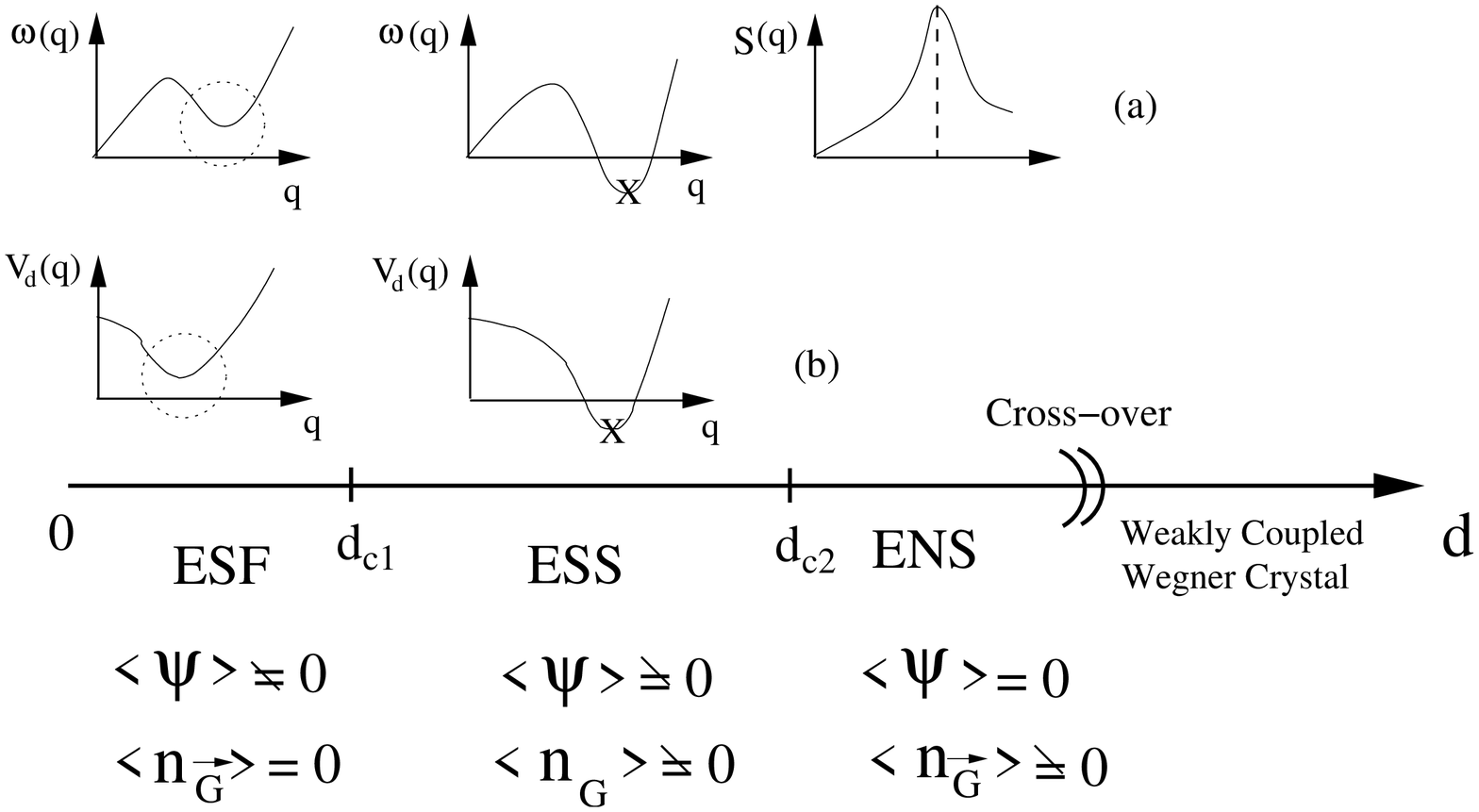}
\caption{ The zero temperature phase diagram in the SEHB as the
distance between the two layers increases. ESF where $ < \psi> \neq
0, < n_{\vec{G} } >=0 $ stands for excitonic superfluid, ESS where $
< \psi> \neq 0, < n_{\vec{G} } > \neq 0 $ stands for exciton
supersolid phase, ENS where $ < \psi> = 0, < n_{\vec{G} } > \neq 0 $
stands for exciton normal solid phase. (a) Energy dispersion
relation $ \omega(q) $ in these phases. (b) the bare dipole-dipole
interaction $ V_{d}(q) $ in these phases. The cross in the ESS means
the negative minimum value of $ V_{d}( q ) $ is replaced by the ESS
state. The order parameters are also shown. Also shown in far right
in (a) is the equal time structure factor $ S(q) $. In fact, the
instability happens before the minimum touches zero. The generic
transition driven by the collapsing of roton minimum is the ESF to
ENS transition. However, as argued in the text, there could be a
window of ESS intervening between the ESF and the ENS. }
 \label{fig1}
\end{figure}



\section{  Existence of ESS intervening between the ESF and ENS. }

    In the section, I point out a mechanism to generate excitonic
    vacancies which lead to a narrow window of ESS intervening between the ESF and the ENS.
     As the distance increases to the critical
     distance $ d_{c1} $, because the lattice constant $ r_{s} a_{B} $ is
     completely fixed by the parameter $ r_{s} $ which is {\em independent of } the distance which drives
     the transition, so the resulting state is likely to have vacancies with density  $ n_{v}(0) $ even at $ T=0 $.
     By contrast, in solid Helium 4, the density is self-determined by
   the pressure $ n= \frac{\partial P}{\partial \mu}|_{T,V} $, so the density and pressure go hand in hand,
   the solid $^{4} He $ is likely to be commensurate.
   We expect that the vacancy-vacancy interaction is also a repulsive dipole-dipole one.
   It is the condensation of these repulsively interacting vacancies at $ T=0 $ which leads to the
   SF mode $ \psi( \vec{x},\tau ) $ inside the {\em in-commensurate} ENS.
   This resulting state is the ESS state where $  \langle \psi \rangle \neq 0, \langle \delta \rho \rangle \neq 0 $ ( Fig.1 ).
   As the distance increases to $ d_{c2} > d_{c1}, n_{v}(0)=0 $, the
   resulting state is a {\em commensurate} ENS whose lattice constant is still {\em locked} at $ r_{s} a_{B} $ (Fig.1).
   As distance increases further, the ENS will crossover to the two
   weakly coupled Wigner crystal.
   Note that because $ V_{d}( \vec{q} \rightarrow 0 ) \rightarrow c
   $, the phonon spectrum is still $ \omega \sim q $ instead of  $ \sim q^{3/2} $ as claimed in \cite{hand}.
   Very similar argument was used in \cite{psdw} to
   conclude that there should be zero-point quantum fluctuations
   generated vacancies in the pseudo-spin density wave ( PSDW ) in BLQH.
   By contrast, due to the absence of the cubic term, the lattice is
   a square lattice. In the PSDW, the vacancies are essentially fermionic holes, so can
   not condensate.
 It is the correlated
   hopping of vacancies in the active and passive layers in the PSDW state  which leads
   to very large and temperature dependent drag consistent with the experimental data.

   The finite temperature phase diagram corresponding to Fig.1 is shown in Fig.2.
   At any finite $ T $, the ESF will only have algebraic long range
   order and will turn into a normal liquid (NL) by a Kosterlitz-Thouless ( KT ) transition. ENS may go
   through a hexatic phase before melting into a NL phase.
   The finite temperature phases and
   phase transitions above the ESS phase in the intermediate distance $ d_{c1} < d < d_{c2} $ is also shown in Fig.2.

\begin{figure}
\includegraphics[width=6cm]{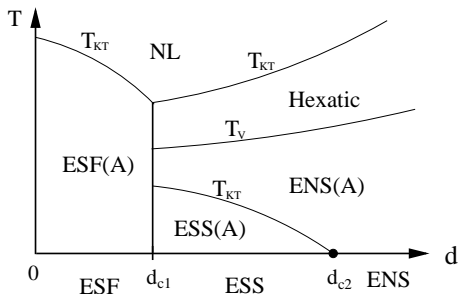}
\caption{ The finite temperature phase diagram in the SEHB as the
distance between the two layers increases. ESF(A) means only
algebraic off-diagonal long range order( ODLRO ), ESS(A) means only
algebraic ODLRO and algebraic translational order.  ENS(A) means
only algebraic translational order. Hexatic phase has the algebraic
orientational order, but no translational order. NL is the normal
liquid phase. $ T_{KT} $ is the KT transition. $ T_{V} $ is the
vector Coulomb gas transition. The dot is the zero temperature
transition from the ENS to the ESS investigated in section 4.}
 \label{fig2}
\end{figure}


\section{ The zero temperature transition from ESS to ENS driven by
the distance. }

     The effective action to describe the ENS to ESS transition at $
     T=0 $ consistent with all the lattice symmetries  and the global
     $ U(1) $ symmetry is:
\begin{eqnarray}
   {\cal L}  & = & \psi^{\dagger} \partial_{\tau} \psi  +
    c_{\alpha \beta} \partial_{\alpha} \psi^{\dagger} \partial_{\beta} \psi + r | \psi |^{2} +
    g  |\psi |^{4}          \nonumber  \\
    & + &  \frac{1}{2} \rho_{n}  ( \partial_{\tau} u_{\alpha} )^{2} +
    \frac{1}{2} \lambda_{\alpha \beta \gamma \delta} u_{\alpha \beta
    } u_{\gamma \delta }     \nonumber  \\
    & + &  a^{0}_{\alpha \beta }  u_{\alpha \beta } \psi^{\dagger} \partial_{\tau} \psi
     + a^{1}_{\alpha \beta } u_{\alpha \beta } | \psi |^{2} + \cdots
\label{quan}
\end{eqnarray}
    where  $ r=d-d_{c2} $ ( Fig. 1),
    $ \rho_{n} $ is the normal density, $ u_{\alpha \beta }= \frac{1}{2}( \partial_{\alpha}
    u_{\beta} + \partial_{\beta} u_{\alpha} ) $ is the linearized  strain
    tensor, $ \lambda_{\alpha \beta \gamma \delta} $ are the bare
    elastic constants dictated by the symmetry of the
    lattice, it has 2  independent elastic constants for a
    triangular lattice where $ \lambda_{\alpha \beta \gamma \delta}=
    \lambda \delta_{\alpha \beta} \delta_{\gamma \delta}
    + \mu ( \delta_{\alpha \gamma } \delta_{\beta \delta}+ \delta_{\alpha \delta} \delta_{\beta \gamma
    } ) $ where $ \lambda $ and $ \mu $ are Lame coefficients. In the ENS state $ r >0, \langle \psi( \vec{x}, \tau
     ) \rangle =0 $, the 2 lattice phonon modes $ \vec{u}( \vec{x}, \tau )  $ become the 2 ordinary
     ones. While inside the ESS state $ r < 0,  \langle \psi( \vec{x}, \tau ) \rangle \neq 0 $.
    If setting all the couplings between $ \psi $ and $ u_{\alpha} $
    vanish, the $ \psi $ sector describes the SF to Mott insulator transition
    in a {\em rigid } underlying two dimensional lattice studied in \cite{boson}.
    Under the Renormalization  group ( RG ) transformation, $
    \tau^{\prime} = \tau/b^{z},  x^{\prime} = x /b $ and $ \psi^{\prime}
    = \psi/Z $. If we choose $ z=2, Z= b^{-d/2} $, the $ g^{\prime}=
    g b^{2-d} $. We also choose  $ u^{\prime}_{\alpha}
    = u_{\alpha}/Z $, then $ \rho^{\prime}_{n}= b^{-2} \rho_{n} $,
    so the lattice phonon kinetic energy term is irrelevant near the QCP.
    It is easy to see $ a^{\prime}_{0} = b^{-d/2-1} a_{0} $, so $
    a_{0} $ is always irrelevant.  $ a^{\prime}_{1} = b^{1-d/2} a_{1}
    $, so both $ g $ and $ a_{1} $'s upper critical dimension is $
    d_{u}=2 $, so we can perform a $ \epsilon=2-d $ expansion in both $ g $ and $ a_{1} $.
    The RG equations are found to be:
\begin{equation}
   \frac{d g}{ d l} = \epsilon g- c g^{2};~~~~\frac{ d a_{1} }{ d
   l}= \frac{\epsilon}{2} a_{1}
\end{equation}
    where $ c= 2m K_{d} \Lambda^{d-2}/ \hbar^{2} $ is the same constant as
    that in the  rigid model \cite{boson}. So $ a_{1} $ is exactly
    marginal without affecting the universality class. This is can
    also be understood as follows: because $
    \rho_{n} $ is irrelevant, so we can simply integrate out $
    u_{\alpha} $ which only leads to a shift of the value of $ g $.
    We conclude that the ENS to ESS transition stays in the same
    universality class of the superfluid to Mott
    insulator transition  at $ d = 2$ in a rigid lattice which has the mean field
    exponents $ z=2, \nu=1/2, \eta=0 $ with logarithmic corrections.
    For example, the superfluid density inside the ESS should scale
    as $ \rho_{s} \sim |d_{c2}-d|^{(d+z-2)\nu }=| d_{c2} -d | $ with a logarithmic correction.

    If neglecting the $ \tau $ dependence by setting
    $ u_{\alpha}( \vec{x}, \tau )= u_{\alpha}( \vec{x} ), \psi_{0}( \vec{x}, \tau ) =   \psi_{0}( \vec{x} ) $,
    then Eqn.\ref{quan} reduces to the classical action.
    For the classical case, $ x^{\prime} = x /b, \psi^{\prime}
    = \psi/Z, u^{\prime}_{\alpha} = u_{\alpha}/Z  $, if we choose $
    Z=b^{(2-d)/2} $, then $  g^{\prime}=
    g b^{4-d}, a^{\prime}_{1} = b^{2-d/2} a_{1} $, so both $ g $ and $ a_{1} $'s upper critical dimension is $
    d_{u}=4 $. So in principle, a $ \epsilon=4-d $ expansion is
    possible for both $ g $ and $ a_{1} $, the putting $ \epsilon=2
    $ for $ d=2 $.   It is known that due to the essential singularity of the KT transition,
    the specific heat exponent of the KT transition $ \alpha = -\infty < 0 $,  the $ a^{1}_{\alpha \beta } $
    coupling is irrelevant, so the ENS to ESS transition remains to be
    Kosterlitz-Thouless (KT) transition at finite temperature. This
    conclusion is consistent with the RG analysis at $ T=0 $ in the
    last paragraph. The quantum critical scalings near the ENS to the ESS
    transition at $ d=d_{c2} $ can be worked out along the similar lines in \cite{scaling}.

 \section{ The low energy excitations in the ESS }

   In this section, we  will study the low energy elementary excitations in the ESS.
   Inside the ESS,  $  \langle \psi_{0}( \vec{x}, \tau ) \rangle = a $, we can
    write $  \psi_{0}( \vec{x}, \tau )= \sqrt{a+ \delta \rho} e^{ i
    \theta (\vec{x}, \tau ) } $ and plug it into the Eqn.\ref{quan}.
     Integrating out the massive magnitude $ \delta \rho $
     fluctuations and simplifying, we get the
     effective action describing the low energy
     modes inside the SS phase:
\begin{eqnarray}
   {\cal L}  & = &  \frac{1}{2} [ \rho_{n}  ( \partial_{\tau} u_{\alpha} )^{2} +
    \lambda_{\alpha \beta \gamma \delta} u_{\alpha \beta
    } u_{\gamma \delta } ]    \nonumber  \\
   & +  &  \frac{1}{2}[ \kappa ( \partial_{\tau} \theta )^{2} +
    \rho^{s}_{\alpha \beta} \partial_{\alpha} \theta \partial_{\beta} \theta ]
    + a_{\alpha \beta }  u_{\alpha \beta } i  \partial_{\tau} \theta
\label{ss1}
\end{eqnarray}
    where  $ \kappa $ is the SF compressibility and $ \rho^{s}_{\alpha \beta} $
    is the SF stiffness which has the same symmetry as $  a^{0}_{\alpha \beta }  $,
    $  a_{\alpha \beta }= a^{0}_{\alpha \beta } + S_{0}
    a^{1}_{\alpha \beta } $ where
    $ S_{0}(\vec{k},\omega ) $ is the bare {\em SF density} correlation function.
    Obviously, the last term is the crucial
    coupling term which couples the lattice phonon modes to the
    SF mode. The factor of $ i $ is important in this coupling.
    By integration by parts, this term can also be written as $ a_{\alpha
    \beta} ( \partial_{\tau} u_{\beta} \partial_{\alpha} \theta
    + \partial_{\tau} u_{\alpha} \partial_{\beta} \theta ) $ which
    has the clear physical meaning of the coupling between the SF
    velocity $ \partial_{\alpha} \theta $ and the velocity of
    the lattice vibration $ \partial_{\tau} u_{\beta} $.
    It is this term which makes
    the low energy modes in
    the SS to have its own characteristics which could be detected by
    experiments. In this section, we neglect the topological
    vortex line excitations in Eqn.\ref{ss1}. In section IX, we will
    discuss these vortex line excitations in detail. In the
    following, we discuss triangular lattice specifically.

    For a triangular lattice, $ \lambda_{\alpha \beta \gamma \delta}=
    \lambda \delta_{\alpha \beta} \delta_{\gamma \delta}
    + \mu ( \delta_{\alpha \gamma } \delta_{\beta \delta}+ \delta_{\alpha \delta} \delta_{\beta \gamma
    } ) $ where $ \lambda $ and $ \nu $ are Lame coefficients, $
    \rho^{s}_{\alpha,\beta}= \rho^{s} \delta_{\alpha,\beta}, a_{\alpha,\beta}= a
    \delta_{\alpha,\beta} $. In $ ( \vec{q}, \omega_{n} ) $ space, the Eqn.\ref{ss1} becomes:
\begin{eqnarray}
    {\cal L}_{is} & = &  \frac{1}{2}[ \rho_{n} \omega^{2}_{n} + ( \lambda+2 \mu
    ) q^{2} ] |u_{l}( \vec{q},\omega_{n} ) |^{2}         \nonumber  \\
     & + & \frac{1}{2} [ \kappa \omega^{2}_{n} + \rho_{s} q^{2} ] |\theta ( \vec{q},\omega_{n} ) |^{2}
                    \nonumber  \\
     & + & 2 a q \omega_{n} u_{l}( -\vec{q}, - \omega_{n} ) \theta ( \vec{q},\omega_{n} )
                    \nonumber  \\
    & + &  \frac{1}{2}[ \rho_{n} \omega^{2}_{n} + \mu q^{2} ] |u_{t}( \vec{q},\omega_{n} ) |^{2}
\label{tri}
\end{eqnarray}
     where $ u_{l}( \vec{q},\omega_{n} )= i q_{i} u_{i}( \vec{q},\omega_{n}
     )/q, u_{t}( \vec{q},\omega_{n} )= i \epsilon_{ij} q_{i} u_{j}( \vec{q},\omega_{n}
     )/q $ are the longitudinal and transverse components of the
     displacement field respectively. Note that Eqn.\ref{tri}
     shows that only the longitudinal component couples to the
     superfluid  $ \theta $ mode, while the transverse component
     is unaffected by the superfluid mode. This is expected, because
     the superfluid mode is a longitudinal density mode itself which dose
     not couple to the transverse modes.

     From Eqn.\ref{tri}, we can identify the
     longitudinal-longitudinal phonon correlation function:
\begin{equation}
  \langle u_{l} u_{l} \rangle= \frac{ \kappa \omega^{2}_{n} + \rho_{s} q^{2} }{
  ( \kappa \omega^{2}_{n} + \rho_{s} q^{2} ) (  \rho_{n} \omega^{2}_{n} + ( \lambda+2 \mu
    ) q^{2} ) + a^{2} q^{2} \omega^{2}_{n} }
\end{equation}
     The $ \langle \theta \theta \rangle $ and $ \langle u_{l} \theta \rangle $ correlation
     functions can be similarly written down. By doing the
     analytical continuation $ i \omega_{n} \rightarrow \omega + i
     \delta $, we can identify the two poles of  all the correlation
     functions at $ \omega^{2}_{\pm}= v^{2}_{\pm} q^{2} $ where the
     two velocities $ v_{\pm} $ is given by:
\begin{widetext}
\begin{equation}
  v^{2}_{\pm}  = [ \kappa( \lambda+2 \mu ) + \rho_{s} \rho_{n} +
   a^{2} \pm  \sqrt{ ( \kappa( \lambda+2 \mu ) + \rho_{s} \rho_{n} +
   a^{2} )^{2}- 4 \kappa \rho_{s} \rho_{n} ( \lambda+2 \mu ) } ]/ 2 \kappa \rho_{n}
\end{equation}
\end{widetext}
    If setting $ a =0 $, then $ c^{2}_{\pm} $ reduces to the longitudinal phonon
    velocity $ v^{2}_{lp}=  ( \lambda + 2 \mu )/ \rho_{n} $ and
    the superfluid velocity $ v^{2}_{s} = \rho_{s}/\kappa $ respectively.
    Of course, the transverse phonon velocity $ v^{2}_{tp}= \mu /
    \rho_{n} $ is untouched. For notation simplicity, in the
    following, we just use $ v_{p} $ for $ v_{lp} $.
    Inside the ESS, due to the very small superfluid density $
    \rho_{s} $, it is expected that $ v_{p} > v_{s} $.
    It is easy to show that $ v_{+} > v_{p}> v_{s} > v_{-} $ and
    $ v^{2}_{+} + v^{2}_{-}  >  v^{2}_{p} + v^{2}_{s} $, but  $ v_{+} v_{-} = v_{p} v_{s} $,
    so $ v_{+} + v_{-} > v_{p} + v_{s} $ ( see
    Fig.3 ).

\begin{figure}
\includegraphics[width=6cm]{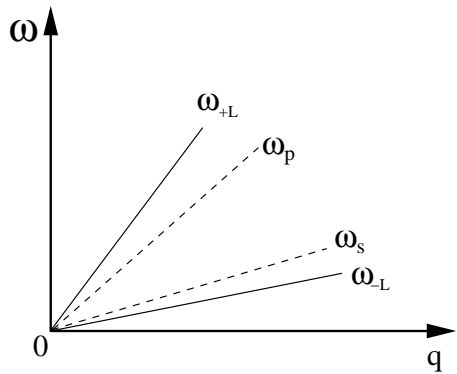}
\caption{ The elementary low energy excitation inside a excitonic
supersolid (ESS). The coupling between the phonon mode $ \omega_{p}
= v_{p}q $ ( the upper dashed line ) and the superfluid mode $
\omega_{s}=v_{s}q $ ( the lower dashed line ) leads to the two new
longitudinal modes $ \omega_{\pm}=v_{\pm} q $ ( solid lines ) in the
ESS. The transverse mode stays the same as that in the ENS and is
not shown.  }
 \label{fig3}
\end{figure}

    In principle, inelastic neutron scattering experiments or acoustic attenuation experiments
    can be used to detect the predicted the low energy excitation spectra in the SS shown in Fig.3.

\section{  Debye-Waller factor in the X-ray scattering from the ESS }

    It is known that due to zero-point quantum motion in any NS at very low temperature, the X-ray
    scattering amplitude $ I (\vec{G}) $ will be diminished by a Debye-Waller (DW) factor
    $ \sim e^{- \frac{1}{2} G^{2} \langle u^{2}_{\alpha}\rangle } $ at $ d=2 $ where $ u_{\alpha} $
    is the lattice phonon modes in Eqn.\ref{ss1}. In Eqn.\ref{ss1},
    if the coupling between the $ \vec{u} $ and $ \theta $ were
    absent, then the DW factor in the SS would be the same as
    that in the NS.
    By taking the ratio $ I_{SS}( \vec{G})/I_{NS}( \vec{G}) $ at a given reciprocal lattice vector
    $ \vec{G} $, then this DW factor drops
    out. Due to the presence of this coupling,  the $ \langle u^{2}_{\alpha} \rangle
    $ in SS is different than that in NS, so the DW factor
    will {\sl not} drop out in the ratio.
    In this section, we will calculate this ratio and see how to take care of this factor
    when comparing with the X-ray scattering data.

    The density order parameter at the reciprocal lattice vector $ \vec{G} $ is
  $ \rho_{ \vec{G} }( \vec{x},\tau )  = e^{ i \vec{G} \cdot \vec{u}( \vec{x},\tau )  } $,
  then $ \langle \rho_{ \vec{G} }( \vec{x},\tau ) \rangle = e^{-\frac{1}{2}
  G_{i}G_{j} \langle u_{i} u_{j} \rangle } $. The Debye-Waller factor:
\begin{equation}
   I( \vec{G} ) =
  | \langle \rho_{ \vec{G} }( \vec{x},\tau ) \rangle |^{2}=  e^{- G_{i}G_{j} \langle u_{i}( \vec{x},\tau ) u_{j}( \vec{x},\tau ) \rangle }
\label{dw}
\end{equation}
  where the phonon-phonon  correlation  function is:
\begin{equation}
   \langle u_{i} u_{j} \rangle =  \langle u_{l}u_{l} \rangle  \hat{q}_{i}\hat{q}_{j} +  \langle u_{t}u_{t} \rangle ( \delta_{ij}- \hat{q}_{i}\hat{q}_{j} )
\label{uu}
\end{equation}
   where $ \hat{q}_{i}\hat{q}_{j}= \frac{ q_{i} q_{j} }{ q^{2}} $.  .

   Then substituting Eqn.\ref{uu} into Eqn.\ref{dw} leads to:
\begin{equation}
  \alpha( \vec{G} ) = I_{SS}( \vec{G} )/I_{NS}(
  \vec{G})= e^{-\frac{1}{2} G^{2}[ \langle u^{2}_{l}( \vec{x},\tau ) \rangle_{SS}- \langle u^{2}_{l}( \vec{x},\tau ) \rangle_{NS} ] }
\end{equation}
  where the transverse mode drops out,  because it stays the same in the SS and in the NS.

  Defining $ ( \Delta u^{2})_{l}( \vec{q}, i \omega_{n}  )=
  \langle | u_{l}( \vec{q}, i \omega_{n}  )|^{2} \rangle_{SS}- \langle | u_{l}( \vec{q}, i \omega_{n}  )|^{2} \rangle_{NS} $,
  $ ( \Delta u^{2} )_{l}( \vec{q} )= \sum_{ i \omega_{n} } ( \Delta u^{2})_{l}(
     \vec{q}, i \omega_{n} ) $  and
  $  ( \Delta u^{2})_{l}= \langle u^{2}_{l}( \vec{x},\tau ) \rangle_{SS}- \langle u^{2}_{l}( \vec{x},\tau ) \rangle_{NS}
  =  \int \frac{ d^{2}q }{ (2 \pi)^{2} } \frac{1}{\beta} \sum_{ i \omega_{n} } ( \Delta u^{2})_{l}(
     \vec{q}, i \omega_{n})  = \int \frac{ d^{2}q }{ (2 \pi)^{2} }( \Delta u^{2} )_{l}( \vec{q} )   $,  it is easy to see:
\begin{widetext}
\begin{equation}
  ( \Delta u^{2})_{l} =  \int \frac{ d^{2}q }{ (2
  \pi)^{2} } \frac{1}{\beta} \sum_{ i \omega_{n} } \frac{- a^{2}
  q^{2} \omega^{2}_{n} }{ [ ( \kappa \omega^{2}_{n} + \rho_{s} q^{2} ) (  \rho_{n} \omega^{2}_{n} + ( \lambda+2 \mu
    ) q^{2} ) + a^{2} q^{2} \omega^{2}_{n} ][  \rho_{n} \omega^{2}_{n} + ( \lambda+2 \mu
    ) q^{2} ]}
\label{long}
\end{equation}
\end{widetext}
    Obviously, $ ( \Delta u^{2})_{l} < 0   $, namely,
    the longitudinal vibration amplitude in SS is {\em smaller} that that
    in NS. Then $ \alpha( \vec{G} )( T=0 )= e^{-\frac{1}{2} G^{2}( \Delta u^{2})_{l}} > 1 $.
    This is expected, because the SS state is the ground
    state at $ T < T_{SS} $, so the longitudinal vibration amplitude
    should be reduced compared to the corresponding NS with the same
    parameters $ \rho_{n}, \lambda, \mu $.

    After evaluating the frequency summation in Eqn.\ref{long}, we get:
\begin{eqnarray}
  ( \Delta u^{2})_{l}(T)  =  \int \frac{ d^{2}q }{ (2
  \pi)^{2} } \frac{1}{ \rho_{n} }[ \frac{ \coth \beta v_{+} q/2 }{ 2
  v_{+} q } - \frac{ \coth \beta v_{p} q/2 }{ 2 v_{p} q }  \nonumber  \\
  -  (\frac{ v^{2}_{s}-v^{2}_{-} }{ v^{2}_{+}-v^{2}_{-} })(
  \frac{ \coth \beta v_{+} q/2 }{ 2
  v_{+} q } - \frac{ \coth \beta v_{-} q/2 }{ 2 v_{-} q } )]
\label{finite}
\end{eqnarray}

   At $ T=0 $, the above equation simplifies to:
\begin{eqnarray}
  ( \Delta u^{2})_{l}( T=0)  =  \int \frac{ d^{2}q }{ (2
  \pi)^{2} } \frac{1}{ \rho_{n} }[ \frac{ 1 }{ 2
  v_{+} q } - \frac{ 1 }{ 2 v_{p} q }  \nonumber  \\
  -  ( \frac{ v^{2}_{s}-v^{2}_{-} }{ v^{2}_{+}-v^{2}_{-} })(
  \frac{ 1 }{ 2 v_{+} q } - \frac{1 }{ 2 v_{-} q } )]  ~~~~~~~~~~~~~~~~~  \nonumber \\
  = - \frac{ ( v_{+}+v_{-}-v_{p}-v_{s}) }{ ( v_{+}+v_{-} ) v_{p} }
  \frac{ \Lambda }{ 4 \pi \rho_{n} } ~~~~~~~~~~~~~~~~ \nonumber \\
   =  - \frac{ a^{2}}{ \kappa \rho_{n} }  \frac{1}{
  (v_{+}+v_{-}+v_{p}+v_{s})( v_{+}+v_{-} ) v_{p} } \frac{ \Lambda }{ 4 \pi \rho_{n}
  } < 0
\label{zero}
\end{eqnarray}
   where  $ \Lambda \sim 1/a $ is the ultra-violet cutoff and
   we have used the fact $ v_{+}+v_{-} > v_{p} + v_{s} $.

   At finite $ T $, in Eqn.\ref{finite}, if taking $ q \rightarrow 0
   $ limit, it is easy to see the integrand identically vanishes.
   So Eqn.\ref{finite} is well defined.

\section{  Density-density correlations }

      The density-density correlation function in the SS is:
\begin{equation}
    \langle \rho_{ \vec{G} }( \vec{x}, t ) \rho^{*}_{ \vec{G} }( \vec{x}^{\prime}, t^{\prime} )
    \rangle= e^{-\frac{1}{2}  G_{i}G_{j}
    \langle ( u_{i}(\vec{x}, t )-u_{i}(\vec{x}^{\prime}, t^{\prime} ))
      ( u_{j}(\vec{x}, t )-u_{j}(\vec{x}^{\prime}, t^{\prime} )) \rangle }
\end{equation}
   where $ t $ is the real time.

    For simplicity, we only evaluate the equal-time correlator
    $  \langle \rho_{ \vec{G} }( \vec{x}, t ) \rho^{*}_{ \vec{G} }( \vec{x}^{\prime}, t
    ) \rangle = \langle \rho_{ \vec{G} }( \vec{x}, \tau ) \rho^{*}_{ \vec{G} }( \vec{x}^{\prime},
    \tau ) \rangle  $ where $ \tau $ is the imaginary time. It is instructive to compare the density order in
    SS with that in a NS by looking at
    the ratio of the density correlation function in the SS over the NS:
\begin{equation}
     \alpha_{\rho}( \vec{x}- \vec{x}^{\prime} )=
     \langle \rho_{ \vec{G} } \rho^{*}_{ \vec{G} }\rangle_{SS}/ \langle \rho_{ \vec{G} } \rho^{*}_{ \vec{G} }\rangle_{NS}
     =  e^{-\frac{1}{4} G^{2} \Delta D_{\rho} ( \vec{x}- \vec{x}^{\prime} ) }
\label{ratiorho}
\end{equation}

    It is easy to find that
\begin{equation}
   \Delta D_{\rho} ( \vec{x}- \vec{x}^{\prime} ) = \int \frac{ d^{2}q }{(2
   \pi)^{2} } ( 2-e^{i \vec{q} \cdot (\vec{x}-\vec{x}^{\prime})}-
                 e^{-i \vec{q} \cdot (\vec{x}-\vec{x}^{\prime})}) ( \Delta u^{2})_{l}( \vec{q} )
\end{equation}
     where $ ( \Delta u^{2})_{l}( \vec{q} ) $ is defined above Eqn.\ref{long} and is the integrand in
     Eqn.\ref{finite}.

     At $ T=0 $, the above equation can be simplified to
\begin{equation}
     \Delta D_{\rho}( \vec{x}- \vec{x}^{\prime} )
     = \frac{ ( v_{+} + v_{-} -v_{p}-v_{s} )}{ (v_{+}+v_{-}) v_{p} }
     \frac{1}{ 2 \pi \rho_{n} } \frac{1}{| \vec{x}- \vec{x}^{\prime} |}
\end{equation}

       So we conclude that $  \alpha_{\rho}( \vec{x}- \vec{x}^{\prime} ) < 1 $, namely,
       the density order in SS is {\em weaker }
     than the NS with the corresponding parameters $ \rho_{n}, \lambda, \mu $.
     This is expected because the density order in the SS is
     weakened by the presence of moving vacancies.

     It is known that there is only algebraic order instead of true solid order at any
     finite $ T $, so the algebraic decay exponent in ESS stays the same as that in ENS.


\section{ Specific heat in the ESS }

     It is easy to see that  at low $ T $, the specific heat in the ENS
\begin{equation}
       C^{NS}= C^{NS}_{lp} + C^{NS}_{tp} + C_{van}
\label{ens}
\end{equation}
     where $ C^{NS}_{lp} =
     \frac{3}{\pi} \zeta(3)  k_{B} ( \frac{k_{B}T}{\hbar v_{lp} })^{2}
     $ is due to the longitudinal phonon mode where $ \zeta(3)=1.202 $  and $ C^{NS}_{tp} =
     \frac{3}{\pi} \zeta(3)  k_{B} ( \frac{k_{B}T}{\hbar v_{tp} })^{2}
     $ is  due to the transverse phonon mode, while $ C_{van} $ is
     from the vacancy contribution. It is still not known how to
     calculate $ C_{van} $ yet.
     The specific heat in the SF $ C^{SF}_{v}= \frac{3}{\pi} \zeta(3)  k_{B} ( \frac{k_{B}T}{\hbar v_{s} })^{2} $
     is due to the SF mode $ \theta $.
     From Eqn.\ref{tri}, we can find the
     specific heat in the ESS:
\begin{equation}
 C^{SS}_{v}= \frac{3}{\pi} \zeta(3)  k_{B} ( \frac{k_{B}T}{\hbar v_{+}
 })^{2} + \frac{3}{\pi} \zeta(3)  k_{B} ( \frac{k_{B}T}{\hbar v_{-}
 })^{2} + C^{tp}
\label{ess}
\end{equation}
    where $ C^{tp} $ stands for the contributions from the
    transverse phonons which are the same as those in the ENS.

 \begin{figure}
\includegraphics[width=6cm]{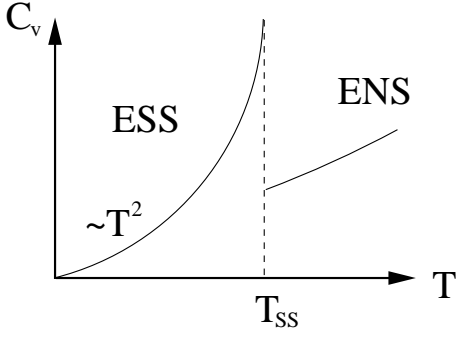}
\caption{ Specific heat in ENS and ESS. The area below $ C_{v}/T $
from $ T=0 $ to $ T=T_{SS} $  is the entropy in the ESS. The excess
entropy is due to the lower branch }
 \label{fig4}
\end{figure}

     It was argued in \cite{qgl}, the critical regime of finite temperature  NS to
     SS transition in Fig.1 is much narrower than the that of SF to
     the NL transition, so there should be a jump in the specific
     heat at $ T=T_{SS} $. Eqn.\ref{ess} shows that at $ T < T_{SS}
     $, the specific heat still takes $  \sim T^{2} $ behavior  and is
     dominated by the $ \omega_{-} $ mode in Fig.2.

 From Eqn.\ref{ess}, it is easy to evaluate the entropy inside the ESS:
\begin{equation}
 S_{SS}(T_{SS})= \frac{3}{2 \pi} \zeta(3)  k_{B} ( \frac{k_{B} T_{SS}
 }{\hbar})^{2}( \frac{1}{ v^{2}_{+} }+ \frac{1}{ v^{2}_{-} } ) + S^{tp}
 \label{enss}
\end{equation}
      where $ S^{tp} $ is the entropy due to the transverse mode.

      From Eqn.\ref{ens} and \ref{ess}, we find that the {\em excess} entropy  due to the vacancy condensation is:
\begin{equation}
 \Delta S = \int^{T_{SS}}_{0} dT  C_{van}/T = \frac{3}{2 \pi} \zeta(3) k_{B} ( \frac{k_{B} T_{SS}
 }{\hbar})^{2}( \frac{1}{ v^{2}_{+} }+ \frac{1}{ v^{2}_{-} }- \frac{1}{ v^{2}_{lp} })
 \label{enss1}
\end{equation}

       Obviously,  $ \Delta S >0 $ is due to the lower branch $
       v_{-} < v_{lp} $. At $ a=0 $, the above equation reduces to $
       \Delta S= \frac{3}{2 \pi} \zeta(3) k_{B} ( \frac{k_{B} T_{SS} }{\hbar v_{s}
       })^{2} $ which is simply dur to the vacancy condensation into ESF.

\section{ Vortices in  the excitonic supersolid }

   In section 5, we studied the low energy excitations shown in the
   Fig.2 by neglecting the  topological vortex line excitations. Here, we will
   study how the vortex line interaction in ESS differ from that in
   the ESF. We can perform a duality transformation on
   Eqn.\ref{ss1} to the vortex  representation:
\begin{equation}
   {\cal L}_{v} = \frac{1}{2 K_{\mu}} ( \epsilon_{\mu \nu \lambda
   } \partial_{\nu} a_{\lambda } - a \partial_{\alpha} u_{\alpha} \delta_{\mu \tau}  )^{2}
    + i 2 \pi a_{\mu } j^{v}_{\mu }
\label{dual}
\end{equation}
    where $ \mu, \nu, \lambda $ stand for space and time, while $ \alpha,\beta $ stand
     for the space components only, $ K_{0}= \kappa, K_{\alpha}= \rho_{s} $ and
     $  a_{\mu } $ is the 3 component  gauge fields and
     $ j^{v}_{\mu }= \frac{1}{2 \pi}  \epsilon_{\mu \nu \lambda
   } \partial_{\nu } \partial_{\lambda } \theta $ is the vortex
     line current due to the topological phase winding in $ \theta $.

    Eqn.\ref{dual} has the gauge invariance $a_{\mu } \rightarrow
    a_{\mu} + \partial_{\mu} \chi $ where $ \chi $ is any field.
    It is most convenient to choose the Coulomb gauge $ \partial_{\alpha} a_{\alpha }=0 $
    to get rid of the longitudinal component, then
    the transverse component is $ a_{t}= i \epsilon_{\alpha \beta} q_{\alpha} a_{\beta} /q $.
     It can be shown that $ |a_{t}|^{2}=| a_{\alpha} |^{2} $.
    Then Eqn.\ref{dual} becomes:
\begin{eqnarray}
 {\cal L}_{v} & = &  \frac{1}{2}[ \rho_{n} \omega^{2}_{n} + ( \lambda+2 \mu
     + a^{2}/\kappa ) q^{2} ] |u_{l}( \vec{q},\omega_{n} ) |^{2}        \nonumber  \\
     & + &  \frac{1}{2} ( q^{2}/\kappa + \omega^{2}_{n}/\rho_{s} ) |a_{t}|^{2}
     + \frac{ q^{2} }{ 2 \rho_{s} } | a_{0} |^{2}
                    \nonumber  \\
     & - &  a q^{2}/\kappa  u_{l}( -\vec{q}, - \omega_{n} ) a_{t}( \vec{q},\omega_{n} )
                    \nonumber  \\
    & + & i 2 \pi j^{v}_{ 0 } a_{0} + i 2 \pi
    j^{v}_{\alpha } a_{\alpha }
\label{prop}
\end{eqnarray}
    where the transverse phonon mode $ u_{t} $ was omitted, because
    it stays the same as in the NS as shown in Eqn.\ref{tri}.

   It is easy to see that only $ a_{t} $ has the dynamics, while $ a_{0} $ is static.
   This is expected, because although $ a_{\mu} $ has 3 non-vanishing components,
   only the transverse component $ a_{t}
   $ has the dynamics which leads to the original gapless superfluid
   mode $ \omega^{2}= v^{2}_{s} q^{2} $. Eqn.\ref{prop} shows that
   it is the longitudinal phonon mode $ u_{l} $ coupled to the
   transverse gauge mode $ a_{t} $.
    The vortex line density is
   $ j^{v}_{ 0 }= \frac{ 1}{2 \pi} \epsilon_{\alpha \beta } \partial_{\alpha} \partial_{\beta} \theta $ and the
    vortex current is $ j^{v}_{ \alpha }= - \frac{ 1}{2 \pi} \epsilon_{\alpha \beta}
    [ \partial_{0},  \partial_{\beta}] \theta $. Integrating out the $ a_{0} $, we
    get the vortex  density-density interaction:
\begin{equation}
     \pi \rho_{s} \int^{\beta}_{0} d \tau \int dx dy j^{v}_{ 0 \alpha
    }( \vec{x}, \tau ) \log |x-y| j^{v}_{ 0 \beta }( \vec{y}, \tau )
\end{equation}
     Namely, the vortex  density- density interaction in SS stays  as $ \sim ln r $ which is the
     same as that in NS ! The vortex energy and the critical
     temperature $ T_{SS} $ in Fig. 2 is solely determined by the
     superfluid density $ \rho_{s} $ only, independent of any other
     parameters in the action Eqn.\ref{tri}.  The critical behaviors of the vortices close to the 2d XY
     transition was just the well known KT transition.

     Integrating out the $ a_{ \alpha  } $,
     we get the vortex  current- current interaction:
\begin{equation}
    (2\pi)^{2}/2 j^{v}_{\alpha} (-\vec{q},-\omega_{n} )
    D_{\alpha \beta} ( \vec{q}, \omega_{n} )  j^{v}_{\beta }(\vec{q},\omega_{n} )
\end{equation}
     where $ D_{\alpha \beta} ( \vec{q}, \omega_{n} )
     =( \delta_{\alpha \beta }- \frac{ q_{\beta} q_{\beta} }{q^{2}} ) D_{t}(\vec{q}, \omega_{n} ) $
     where  $ D_{t}(\vec{q}, \omega_{n}
     ) $ is the $ a_{t} $ propagator. Defining
      $  \Delta D_{t}(\vec{q}, \omega_{n}) = D^{SS}_{t}(\vec{q}, \omega_{n})-D^{SF}_{t}(\vec{q}, \omega_{n}) $
      as the difference between the $a_{t} $ propagator in the ESS and the ESF,
      then from Eqn.\ref{prop}, we can get:
\begin{equation}
  \Delta D_{t} = \frac{ a^{2} \rho^{2}_{s} q^{4} }{ \kappa \rho_{n}
  ( \omega^{2}_{n} + v^{2}_{+} q^{2} ) ( \omega^{2}_{n} + v^{2}_{-} q^{2} )( \omega^{2}_{n} + v^{2}_{s} q^{2} ) }
\end{equation}

   For simplicity, we just give the expression for the equal time:
\begin{widetext}
\begin{equation}
  \Delta D_{t}( \vec{x}-\vec{x}^{\prime},\tau=0 )
   = \frac{ a^{2} \rho^{2}_{s} }{ 4 \pi \kappa^{2} \rho^{2}_{n} } \frac{
   v_{+} + v_{-} +v_{s} }{ ( v_{+} + v_{-} )( v_{s}+ v_{+})(
   v_{s}+v_{-}) v_{+} v_{-} v_{s}  } \frac{1}{ |  \vec{x}-\vec{x}^{\prime} | }
\end{equation}
\end{widetext}
   Namely, the vortex current-current interaction in SS is stronger
   than that in the SF  with the same parameters $ \kappa, \rho_{s} $ !

\section{ Detection of the ESS by its photon emission patterns. }

     In the previous sections, we worked out the fundamental properties of the
     ESS and various experimental signatures of the ESS. These
     experiments can be performed easily in the gate voltage
     generated exciton systems in \cite{field1,field2}, but it is very hard to perform for
     the photon pumping generated excitons systems in
     \cite{butov,snoke,bell} due to the short life time of these
     kind of excitons. One important question is how to detect the existence of the metastable ESS in Fig.1
     if it indeed exists ? In a recent preprint \cite{si}, the authors found that
     when the angle resolved power spectrum (ARPS) from the ESF in the Fig.1 always takes
     a {\em macroscopic} superradiance form which is proportional to $ N^{2}
     $.  The concept of super-radiance  was first proposed by Dicke in
     1954 for $ N $ two level atomic atoms confined into a small
     volume $ V $ which is much smaller than the wavelength
     of the emitted photon \cite{dicke}. The macroscopic superradiance from the
     ESF even holds in the thermodynamic limit where $ N \rightarrow
     \infty, V \rightarrow \infty $, but keep $ N/V $ at finite.
     This is due to the macroscopic phase coherence of all the $ N $ excitons
     in the whole volume $ V $, so it a completely different
     mechanism than that of Dicke in conventional quantum optics.
     This fact can be used to detect the ESS easily in a ARPS experiment for the following reason:
     the power spectrum from the ENS in Fig.1 is just a normal
     radiance, so proportional to $ N_{NS} $, however, that from the
     ESF in Fig.1 is a macroscopic super-radiance, so proportional to $ N^{2}_{vac}
     $.  Because the ENS is a uniform phase consisting of a normal solid component and a superfluid component,
     the ARPS at any given in-plane momentum $ \vec{k} $ from the ESS is:
\begin{equation}
      S_{ESS}( \vec{k} ) \propto N^{2}_{vac} + N_{NS}
\end{equation}
     which takes a Lorentzian form with the photon energy
     $ \omega_{k}= v_{g} \sqrt{ \vec{k}^{2} + k^{2}_{z} } $ centering around
     the exciton gap $ E_g \sim 1.545 eV $\cite{si}.

     For a simple estimate, the superfluid component is only a small
     $ \sim 10^{-2} $ component of the whole system, so $
     N_{vac}/N_{ NS} \sim 10^{2} $, then $ S_{ESS}( \vec{k} )  \propto N_{NS} ( 10^{-4} N_{NS}
     + 1) $. Although $ 10^{-4} $ is a very tiny number, but it
     times a huge number $ N_{NS} \sim 10^{10} $ for a $ 1 cm^{2}  $
     sample. This leads to the fantastic fact that even the ESF component is
     just a very small component of the ESS, it {\em dominates} the ARPS due to its macroscopic super-radiance.
     So as distance changes in Fig.1, the ARPS along any direction
     should distinguish the three phases ESF, ESS and ENS easily and
     without any ambiguity.

     It was shown in \cite{si}, due to the symmetry breaking in the
     ESF, the emitted photon along the normal direction is a
     coherent state. The photon number is proportional to the
     condensation fraction $ N_{0} $. It was also shown that due to the non-vanishing
     anomalous Green function in the ESF state in Fig.1a, the emitted photons along
     all tilted directions are always in a two mode squeezed
     state between $ \vec{k} $ and $ -\vec{k} $ even off the
     resonance.  The squeezing ratio is proportional to the number of
     excitons $ N $ in the EHBL. All the results are robust and independent of any microscopic
     details. The detection of the coherent state along the normal direction and
     the squeezed state along all the tilted directions requires phase
     sensitive homodyne experiment which is essentially a phase
     interference experiment between the emitted light and a local
     reference oscillator. When increasing the distance as shown in
     the Fig.1, the coherent photons along the normal direction will
     be much reduced from the ESF to the ESS by $ 10^{-2} $ and completely
     disappear inside the ENS. The two modes squeezing ratio will also be much
     reduced from the ESF to ENS by $ 10^{-2} $ and completely
     disappear inside the ENS. However, due to the tiny superfluid component inside the ENS,
     the phase sensitive homodyne experiment to
     measure the two mode squeezing state inside the ENS is much harder to perform
     than the angle resolved power spectrum experiment.

     In principle, the emitted photons from the ESS will form a photonic band
     structure reflecting the periodic triangular lattice structure of the
     ESS. However as shown in \cite{si}, the maximum
     in-plane wavevector of the emitted photons around energy $ E_g \sim 1.545 eV $ is at least 2 order of
     magnitude smaller than the size of the Brillouin Zone of the ESS, so the
     effect of the underlying lattices is negligibly small.

\section{ Conclusions. }

     We  studied  phases and quantum phase transitions in the dilute limit ( $ r_{s} \gg 1 $ ) as distance is increased.
     When the distance is sufficiently small, the system is in the
     Excitonic superfluid state. When the distance is sufficiently
     large, it is in the weakly coupled Wigner crystal state.
     We  argued that there is a Excitonic supersolid state intervening
     between the two phases at some intermediate distance range $
     \sqrt{r_{s}} < d/a_{B} < r_{s} $. We derive a quantum
     Ginsburg-Landau action to describe the ESF to the ESS
     transition driven by the collapsing of the roton minimum.  In general, there
     should be zero-point quantum fluctuations generated vacancies
     whose condensation lead to the formation of superfluid mode
     inside the ESS. We then derive the quantum GL action to
     describe the transition from the ENS at the large distance to
     the ESS at the intermediate distance. By RG analysis, we showed that the coupling to the quantum fluctuations of the
     underlying lattice is marginal, so the ENS to
     the ESS transition is in the same universality class as
     superfluid to Mott insulator transition in a rigid lattice, therefore has exact
     exponents $ z=2, \nu=1/2, \eta=0 $  with logarithmic
     corrections. We showed that the ESS is a triangular lattice. We then study the elementary excitations inside
     the ESS.  We found that the coupling to quantum lattice phonons is very
     important inside the ESS  and leads to two longitudinal supersolidon modes $
     \omega_{\pm}=v_{\pm} q $ shown in Fig.3. The transverse modes
     in the ESS stays the same as those in the ENS. Detecting the two
     longitudinal modes, especially, the lower branch $ \omega_{-} $
     mode by neutron scattering or acoustic wave attention
     experiments is a smoking gun experiment to prove or disprove
     the existence of the ESS in the intermediate distance regime. Then we calculated the experimental
     signature of the two modes. We found that the longitudinal
     vibration in the ESS is smaller than that in the ENS ( with the same corresponding solid parameters ),
     so the Debye-Waller
     factor at a given reciprocal lattice vector is larger than
     that in the ENS. The density-density correlation function in the
     ESS is weaker than that in the ENS.  The specific heat
     in the SS is still given by the sum from the transverse phonons and the two longitudinal
     phonons and still shows $ \sim T^{2} $ behaviors.
     The contribution from longitudinal part is dominated by the lower supersolidon branch in Fig.3.
     By going the to the dual
     vortex representation, we found the vortex
     density-density interaction in ESS stays the same as that in the
     ESF ( with the same corresponding superfluid parameters ), so
     the vortex  energy and the corresponding ESS(A) to ENS(A) transition
     temperature  $ T_{KT} $ in Fig.2 is solely determined by the superfluid density and
     independent of any other parameters. The vortex current-current
     interaction in stronger than that in the SF.
     In principle, all these predictions can be
     tested by experimental techniques such as X-ray scattering,
     neutron scattering, acoustic wave attenuations and heat
     capacity.

     For the excitons generated by photon pumping, the ESS is a meta-stable non-equilibrium
     state, it will decay eventually by emitting photons.
     The angle resolved power spectrum ( ARPS ) $ S_{ESS}(\vec{k} ) $ at any given in-plane momentum
     $ \vec{k} $ is dominated by the macroscopic super-radiance from
     its superfluid component, even the superfluid component is
     just a very small percentage of the the whole system.
     In the ESF, ESS and ENS phases in the Fig.1,
     the ARPS shows the following sequence: $ S_{ESF} \sim N^{2} \gg S_{ESS} \sim N^{2}_{vac} \gg
     S_{ENS} \sim N $. This sequence can be measured by a ARPS along
     any direction easily and without any ambiguity.


     The research at KITP-C is supported by
     the Project of Knowledge Innovation Program (PKIP) of Chinese Academy of Sciences.


\end{document}